\documentclass[12pt]{article}
\usepackage{amsfonts}
\usepackage{amssymb}
\usepackage{amsmath}
\usepackage{graphicx}

\begin{document}

\title{Knots in $SU\left( M|N\right) $ Chern-Simons Field Theory}
\author{Xin LIU \  \\
School of Mathematics and Statistics, University of Sydney\\
NSW 2006, Australia\ \\
liuxin@maths.usyd.edu.au}
\date{ }
\maketitle

\begin{abstract}
Knots in the Chern-Simons field theory with
Lie super gauge group $SU\left( M|N\right) $ are studied, and the $%
S_{L}\left( \alpha ,\beta ,z\right) $ polynomial invariant with
skein relations are obtained under the fundamental representation
of $\mathfrak{su}\left( M|N\right) $.

\bigskip


\end{abstract}

\section{Introduction}

Chern-Simons (CS) theories are Schwarz-type topological field
theories --- a CS action is both gauge invariant and generally
covariant, and a quantum CS theory has general variance in the
BRST formalism under the Landau gauge although a metric enters the
gauge-fixing term \cite {GuadagniniBook}. CS theories were first
introduced into physics in the study of quantum anomaly of gauge
symmetries by Jackiw et al. \cite{JackiwWZW}. Witten pointed out
\cite{WittenCMP1989knots} that CS theories provide a field
theoretical origin for polynomial invariants of links in knot
theory. Different Lie gauge groups of the CS theories and
different algebraic representations of the gauge groups lead to
different link invariants
\cite{WittenCMP1989knots,GuadagniniNPB1990,Kauffmanbook}.
Perturbative expansions of correlation functions of Wilson loops
in CS theories present Vassiliev invariants
\cite{Bar-Natan,BirmanLin,Labastida}. Recent developments include
the applications of CS theories in topological string
theory \cite{Marino2005RMP} and the $\left( 2+1\right) $%
-dimensional quantum gravity \cite{2+1quantumgravity}.

Super symmetries have found realizations in various physical
systems \cite{LieSuperalgebra}. Representation theories for
Lie superalgebras have been developed by many authors \cite%
{Scheunert,RBZetalAlgeRep}. Link invariants have been obtained
from quantum super group invariants by Gould, Bracken, Zhang,
Links, Kauffman, et al. from the algebraic point of view
\cite{RBZlinkInv}, including the HOMFLY polynomial from the
$U_{q}\left( \mathfrak{su}\left( M|N\right) \right) $ invariants
$\left( M\neq N\right) $, the Kauffman polynomial from the
$U_{q}\left( \mathfrak{osp}\left( M|2N\right) \right) $
invariants, and the Alexander-Conway polynomial from the
$U_{q}\left( \mathfrak{gl}\left( N|N\right) \right) $ invariants.

In this paper we will use the field theoretical point of view to
study knots in the CS field theory with super gauge group
$SU\left( M|N\right) ,\ M\neq N$ \cite{CSsuperalge,LiuX}. Under
the fundamental representation of the superalgebra
$\mathfrak{su}\left( M|N\right) $, a correlation function of
Wilson loop operators will be studied and the $S_{L}\left(
\alpha ,\beta ,z\right) $ link polynomial be obtained \cite%
{GuadagniniNPB1990}. One will discuss the relationships between the $%
S_{L}\left( \alpha ,\beta ,z\right) $ polynomial and the HOMFLY and Jones
polynomials, and show that the CS theory with super group $%
SU\left( N+2|N\right) $ has the Jones polynomial invariant. This
is different from the situation of the CS theory with normal Lie
group $ SU\left( N\right) $ --- under the fundamental
representation, only the $SU\left( 2\right) $ CS theory has the
Jones polynomial.

This paper is arranged as follows. In Section 2, the notation of
Lie superalgebra $\mathfrak{su}\left( M|N\right) $ under the
fundamental representation is given. In Section 3, path variation
within correlation functions of Wilson loops in the CS theory is
rigorously studied. In Section 4, the variation of correlation
functions obtained in Section 3 is formally discussed with respect
to different link configurations, without integrating out the path
integrals. From the formal analysis the $S_{L}\left( \alpha ,\beta
,z\right) $ polynomial with skein relations is obtained, and its
relationships to other knot polynomials are discussed. The paper
is summarized in Section 5.

\section{Notation and Preliminary}

Let us fix the notation of the superalgebra
$\mathfrak{su}\left(M|N\right)$ first. Consider the elements
$\left\{ \left. \hat{e} _{ab}\right\vert a,b=1,\cdots ,M+N,\ M\neq
N\right\} $ satisfying the
following super commutation relations \cite%
{KacAnnMath1977,dictionary,glNNdef-NPB1994,WLYangJMP}:%
\begin{equation}
\left[ \hat{e}_{ab},\hat{e}_{cd}\right] =\hat{e}_{ad}\delta _{bc}-\left(
-1\right) ^{\left( \left[ a\right] +\left[ b\right] \right) \left( \left[ c%
\right] +\left[ d\right] \right) }\hat{e}_{cb}\delta _{da}.  \label{2Ecomm}
\end{equation}%
Here the $\mathbb{Z}_{2}$-grading is given by $\left[ \hat{e}_{ab}\right] =%
\left[ a\right] +\left[ b\right] $ with $\left[ 1\right] =\cdots =\left[ M%
\right] =0$ and $\left[ M+1\right] =\cdots =\left[ M+N\right] =1$. In the
fundamental representation $\hat{e}_{ab}$ is realized by%
\begin{equation}
\hat{e}_{ab}=e_{ab}-\frac{\delta _{ab}\left( -1\right) ^{\left[ a\right] }}{%
M-N}I,  \label{defeabhat}
\end{equation}%
where $e_{ab}$ is the $\left( M+N\right) \times \left( M+N\right) $ matrix
unit with entry $1$ at the position $\left( a,b\right) $ and $0$ elsewhere. $%
\hat{e}_{ab}$ satisfies the traceless requirement $Str\left( \hat{e}%
_{ab}\right) =0,$ where $Str\left( X\right) $ is the supertrace of
the representation matrix of $X\in \mathfrak{g}$, $Str\left(
X\right) =\sum_{i}\left( -1\right) ^{\left[ i\right] }X_{ii}$, $i$
denoting the entry
indices. The $\hat{e}_{ab}$'s have the identity $\sum_{a=1}^{M+N}\hat{e}%
_{aa}=0$. The $\left( M+N\right) ^{2}-1$ generators of the supergroup $%
SU\left( M|N\right) $, denoted by $\left\{ \hat{E}_{ab},\hat{F}_{ab},\hat{H}%
_{cc}\right\} $, can be constructed in terms of $\hat{e}_{ab}$:%
\begin{equation}
\begin{array}{ll}
\hat{E}_{ab}=\frac{i}{2}\left( \hat{e}_{ab}-\hat{e}_{ba}\right) ,\ \hat{F}%
_{ab}=\frac{1}{2}\left( \hat{e}_{ab}+\hat{e}_{ba}\right) , & \ \ \
a,b=1,\cdots ,M+N,\ a\neq b; \\
\hat{H}_{cc}=\sum_{l=1}^{c}l\left( \hat{e}_{ll}-\hat{e}_{l+1,l+1}\right) , &
\ \ \ c=1,\cdots ,M+N-1,%
\end{array}
\label{SUMNgen}
\end{equation}%
where no summation for repeating $c,l$. The $\hat{E}_{ab},\hat{F}_{ab}$ and $%
\hat{H}_{cc}$ satisfy the properties of tracelessness and
unitarity: $Str\left( \hat{E} _{ab}\right) =Str\left(
\hat{F}_{ab}\right) =Str\left( \hat{H}_{cc}\right)
=0$; $\left( \hat{E}_{ab}\right) ^{\dag }=\hat{E}_{ab}$, $%
\left( \hat{F}_{ab}\right) ^{\dag }=\hat{F}_{ab}$ and $\left( \hat{H}%
_{cc}\right) ^{\dag }=\hat{H}_{cc}$. The $\hat{E}_{ab}$ and
$\hat{F}_{ab}$ play the role of the raising/lowering generators,
and $\hat{H}_{cc}$ the elements of the Cartan subalgebra of
$\mathfrak{su}\left( M|N\right) $. Hereinafter for convenience one
uses the basis $\left\{ \hat{e}_{ab},\ a\neq b;\ \ \hat{e}_{cc},\
c=1,\cdots ,M+N-1\right\} $.

We begin the study of the knots in a CS field theory by
considering the correlation function of Wilson loops under the
fundamental
representation of $\mathfrak{su}\left( M|N\right) $ \cite%
{WittenCMP1989knots,GuadagniniNPB1990,Kauffmanbook}%
\begin{equation}
\left\langle W\left( L\right) \right\rangle =\left\langle
StrPe^{i\oint_{L}A_{\mu }\left( x\right) dx^{\mu }}\right\rangle
=Z^{-1}StrP\int \mathcal{D}Ae^{iS}e^{i\oint_{L}A_{\mu }\left( x\right)
dx^{\mu }},  \label{Eq1}
\end{equation}%
where $Z=\int \mathcal{D}Ae^{iS}$ the normalization factor. $L$
denotes the integration loop and $P$ the proper product. $S$ is
the non-Abelian CS action,%
\begin{equation}
S=\frac{k}{4\pi }\int_{\mathbb{R}^{3}}d^{3}x\epsilon ^{\mu \nu \rho
}Str\left( A_{\mu }\partial _{\nu }A_{\rho }+\frac{2}{3}A_{\mu }A_{\nu
}A_{\rho }\right) ,  \label{CSorigdef1}
\end{equation}%
$k$ being an integer valued constant. $A_{\mu }$ is the $SU(M|N)$ gauge
potential, $A_{\mu }=A_{\mu }^{ab}\hat{e}_{ab}$. The gauge field tensor $%
F_{\mu \nu }$ is induced by $A_{\mu }:$%
\begin{equation}
F_{\mu \nu }=F_{\mu \nu }^{ab}\hat{e}_{ab},\ \ \ \ \ \ \ \ \ F_{\mu \nu
}^{ab}=\partial _{\mu }A_{\nu }^{ab}-\partial _{\nu }A_{\mu }^{ab}-\left(
-1\right) ^{\left( \left[ a\right] +\left[ c\right] \right) \left( \left[ c%
\right] +\left[ b\right] \right) }\left( A_{\mu }^{ac}A_{\nu
}^{cb}-A_{\nu }^{ac}A_{\mu }^{cb}\right) .  \label{Fuv2}
\end{equation}%
The grading $\left[ A_{\mu }\right] =\left[ F_{\mu \nu }\right] =\left[ S%
\right] =\mathrm{even}$.

The gauge invariance of the phase of the action, $e^{iS}$, needs
more discussion. The gauge transformations of $A_{\mu }$ and
$F_{\mu \nu }$ are $A_{\mu }\longrightarrow \Omega A_{\mu }\Omega
^{-1}+\partial _{\mu }\Omega \Omega ^{-1}$ and $F_{\mu \nu
}\longrightarrow \Omega F_{\mu \nu }\Omega ^{-1},$ with $\Omega $
denoting a group $G$ transformation. It is known that if $G$ is a
normal Lie group the action $S$ transforms as
\begin{equation}
S\longrightarrow S+\frac{k}{4\pi }\int_{\mathbb{R}^{3}}d^{3}x\partial _{\mu
}j^{\mu }+2\pi k\frac{1}{24\pi ^{2}}\int_{\mathbb{R}^{3}}d^{3}x\epsilon
^{\mu \nu \rho }Str\left[ a_{\mu }a_{\nu }a_{\rho }\right] ,
\label{CS-WZW-1}
\end{equation}%
where $a_{\mu }=\Omega ^{-1}\partial _{\mu }\Omega $ and $j^{\mu
}=\epsilon ^{\mu \nu \rho }Str\left( A_{\nu }a_{\rho }\right)$.
The second term in (\ref {CS-WZW-1}) is a total divergence which
has no contribution to the action as
$j^{\mu }$ vanishes at infinity. The third term, marked as $S_{\mathrm{WZW}%
} $, is a Wess-Zumino-Witten (WZW) term. Jackiw, Cronstr\"{o}m,
Mickelsson, et al. \cite{JackiwWZW,CronstromMickelsson} examined
this term for an arbitrary non-Abelian Lie group $G$. They pointed
out that when $\Omega$ satisfies the regular condition --- $\Omega
$ tends to a definite limit
at infinity, $\lim_{\mathbf{x}\rightarrow \infty }\Omega \left( \mathbf{x}%
\right) =I$ --- the WZW term is a total differential%
\begin{equation}
S_{\mathrm{WZW}}=2\pi k\frac{1}{24\pi ^{2}}\int_{\mathbb{R}^{3}}dx^{\mu
}\partial _{\mu }\left[ \Theta _{\nu \rho }dx^{\nu }\wedge dx^{\rho }\right]
=2\pi k\frac{1}{\pi ^{2}}\int_{\mathbb{R}^{3}}d\Theta ,  \label{CS-WZW-4}
\end{equation}%
where $\Theta $ is a $2$-form constructed by $\Omega $, and
$d\Theta $ serves as a volume element \cite{CronstromMickelsson}.
Since the regular
condition implies the compactification $\mathbb{R}^{3}\longrightarrow S^{3}$%
, Eq.(\ref{CS-WZW-4}) becomes $S_{\mathrm{WZW}}=2\pi k\frac{1}{\pi ^{2}}%
\int_{S^{3}}d\Theta $, which gives the degree of the homotopy mapping $%
\Omega :S^{3}\rightarrow G$ when $G$ is compact. Hence for a
compact group $G$ one has $S_{\mathrm{WZW}}=2\pi kw\left( \Omega
\right)$, and the action transforms as $S\rightarrow S+2\pi
kw\left( \Omega \right) $, where $w\left( \Omega \right) $ is the
so-called winding number, $w\left( \Omega \right) \in \pi
_{3}\left[ SU\left( M|N\right) \right] =\mathbb{Z}$. In this
paper, the gauge group is the super group $SU\left( M|N\right) $;
a point needs clarification is whether the WZW term is able to be
written as a total differential. This problem is being studied by
us at present and will be discussed in our further papers.

Under the fundamental representation (\ref{defeabhat}) the $\hat{e}_{ab}$
has the following supertraces%
\begin{eqnarray}
Str\left( \hat{e}_{ab}\hat{e}_{cd}\right) &=&\left( -1\right) ^{\left[ a%
\right] }\delta _{ad}\delta _{bc}-\frac{\left( -1\right) ^{\left[ a\right] +%
\left[ c\right] }\delta _{ab}\delta _{cd}}{M-N},  \label{str2E} \\
Str\left( \hat{e}_{ab}\hat{e}_{cd}\hat{e}_{ef}\right) &=&\left( -1\right) ^{%
\left[ a\right] }\delta _{af}\delta _{bc}\delta _{de}-\left( -1\right) ^{%
\left[ a\right] +\left[ c\right] }\frac{\delta _{ab}\delta _{cf}\delta _{de}%
}{M-N}-\left( -1\right) ^{\left[ c\right] +\left[ f\right] }\frac{\delta
_{cd}\delta _{af}\delta _{be}}{M-N}  \notag \\
&&-\left( -1\right) ^{\left[ f\right] +\left[ a\right] }\frac{\delta
_{ef}\delta _{ad}\delta _{bc}}{M-N}+2\left( -1\right) ^{\left[ a\right] +%
\left[ c\right] +\left[ e\right] }\frac{\delta _{ab}\delta _{cd}\delta _{ef}%
}{\left( M-N\right) ^{2}}.  \label{str3E}
\end{eqnarray}%
In terms of (\ref{str2E}) and (\ref{str3E}) the component form of
the $SU\left( M|N\right) $ CS action reads%
\begin{eqnarray}
S &=&\frac{k}{4\pi }\int d^{3}x\epsilon ^{\mu \nu \rho }\left( -1\right) ^{%
\left[ b\right] }  \notag \\
&&\left[ A_{\mu }^{ab}\partial _{\nu }A_{\rho }^{ba}+\frac{2}{3}\left(
-1\right) ^{\left[ c\right] +\left[ a\right] \left[ b\right] +\left[ b\right]
\left[ c\right] +\left[ c\right] \left[ a\right] }A_{\mu }^{ab}A_{\nu
}^{bc}A_{\rho }^{ca}-\left( -1\right) ^{\left[ a\right] }\frac{A_{\mu
}^{aa}\partial _{\nu }A_{\rho }^{bb}}{M-N}\right.  \notag \\
&&\left. -\frac{2}{3}\left( -1\right) ^{\left[ a\right] }\frac{A_{\mu
}^{aa}A_{\nu }^{cb}A_{\rho }^{bc}}{M-N}+\frac{4}{3}\left( -1\right) ^{\left[
a\right] +\left[ c\right] }\frac{A_{\mu }^{aa}A_{\nu }^{bb}A_{\rho }^{cc}}{%
\left( M-N\right) ^{2}}\right] .  \label{S3}
\end{eqnarray}%
It can be proved that $S$ has an important property \cite%
{Kauffmanbook,GuadagniniNPB1990,GuadagniniBook}%
\begin{equation}
\frac{2\pi }{k}\epsilon ^{\mu \nu \rho }\left( -1\right) ^{\left[ b\right] }%
\frac{\partial S}{\partial A_{\rho }^{ab}\left( x\right) }\hat{e}%
_{ba}=F_{\mu \nu }^{ba}\left( x\right) \hat{e}_{ba}.
\label{CSgoodprop}
\end{equation}%
This gives the equation of motion of a pure gauge:
$\frac{2\pi}{k}\frac{\delta S}{\delta A}=F=0$, which is the same
as the commonly known equation of motion in the CS theories with
normal Lie gauge groups. Eq.(\ref{CSgoodprop}) will be crucial in
following sections for derivation of the skein relations of knots
in the CS theory with $SU\left( M|N\right)$ gauge group.

\section{Variation of Correlation Function}

In this section correlation functions of Wilson loops will be
studied, with emphasis placed on variation of integration paths
and the induced changes of the correlation functions.

Consider two knots which are almost the same except at one double-point $%
x_{0}$, as illustrated by Figure 1.
\begin{figure}[th]
\centering \includegraphics[scale=0.9]{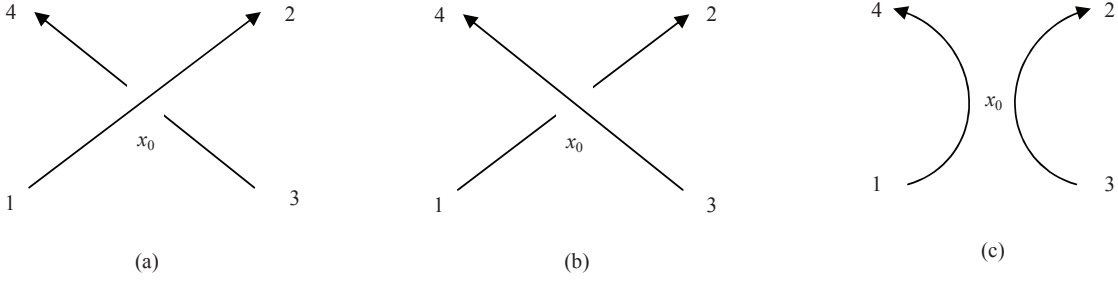}
\caption{Overcrossing, Undercrossing and Non-Crossing: (a) $L_{+}$; (b) $L_{-}$; (c) $L_{0}$.}
\end{figure}

\noindent Here $1,2,3,4$ are the abbreviations for the points $%
x_{1},x_{2},x_{3},x_{4}$. Denote the knot in Figure 1(a) as
$L_{+}$ and that in Figure 1(b) as $L_{-}$. Figure 1(c) shows the
non-crossing situation. Let $U\left( 1,2\right) $ [resp. $U\left(
3,4\right) $] be the propagation process along the segment $\left(
1\rightarrow 2\right) $ [resp. $\left( 3\rightarrow 4\right) $].
For convenience denote the $U\left( 1,2\right) $ in Figure 1(a) as
$U_{+}\left( 1,2\right) $, and that in Figure 1(b) as $U_{-}\left(
1,2\right) $. In both Figures 1(a) and 1(b), the process $\left(
1\rightarrow 2\right) $ is prior to $\left( 3\rightarrow 4\right)
$ in the sense of proper order. In following we will discuss the
difference between the overcrossing $L_{+}$ and undercrossing
$L_{-}$, by fixing the segment $\left( 3\rightarrow 4\right) $ and
moving the segment $\left( 1\rightarrow 2\right) $ from back to
front.

Let $\left\langle W\left( L_{+}\right) \right\rangle \ $and
$\left\langle W\left( L_{-}\right) \right\rangle $ be the
respective correlation functions of $L_{+}$ and $L_{-}$. Each of
them can be written as a series
of propagation processes in proper order:%
\begin{equation}
\left\langle W\left( L_{\pm }\right) \right\rangle =\left\langle Str\left[
\cdots U_{\pm }\left( 1,2\right) \cdots U\left( 3,4\right) \cdots \right]
\right\rangle ,  \label{Eq2}
\end{equation}%
where the propagators are realized by%
\begin{equation}
U_{\pm }\left( 1,2\right) =\left. e^{i\int_{1}^{2}A_{\mu }\left( x\right)
dx^{\mu }}\right\vert _{L_{\pm }},\ \ \ \ \ \ \ \ \ U\left( 3,4\right)
=e^{i\int_{3}^{4}A_{\mu }\left( x\right) dx^{\mu }},
\end{equation}%
the grading of $U_{\pm }\left( 1,2\right) $ and $U\left(
3,4\right) $ being even. The difference between the correlation
functions of $L_{+}$ and $L_{-}$ is
\begin{equation}
\left\langle W\left( L_{+}\right) \right\rangle -\left\langle W\left(
L_{-}\right) \right\rangle =\left\langle Str\left( \cdots \left[ U_{+}\left(
1,2\right) -U_{-}\left( 1,2\right) \right] \cdots U\left( 3,4\right) \cdots
\right) \right\rangle .  \label{W+-pathint}
\end{equation}%
The path variation $L_{-}\rightarrow L_{+},$ given by $\left[
U_{+}\left( 1,2\right) -U_{-}\left( 1,2\right) \right] $ in
(\ref{W+-pathint}), is stereoscopically illustrated in Figure 2,
where the segment $\left( 1\rightarrow 2\right) $ in $L_{-}$
corresponds to the path $\overline{\mathrm{1ACDB2}}$, and that in
$L_{+}$ to $\overline{\mathrm{1AEFB2}}$.
\begin{figure}[th]
\centering \includegraphics[scale=0.7]{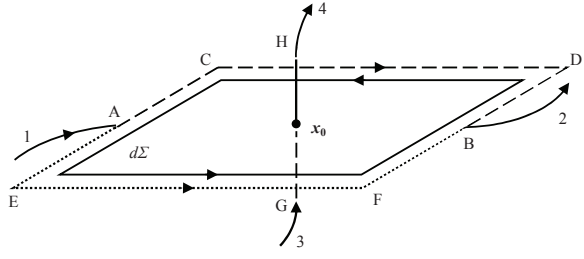}
\caption{$3$-Dimensional Geometric Illustration of Path Variation}
\end{figure}
Then%
\begin{equation}
U_{+}\left( 1,2\right) -U_{-}\left( 1,2\right) =U\left( 1,A\right) \left(
i\int_{\overline{\mathrm{AEFB}}}A_{\mu }\left( x\right) dx^{\mu }-i\int_{%
\overline{\mathrm{ACDB}}}A_{\mu }\left( x\right) dx^{\mu }\right) U\left(
B,2\right) ,  \label{Eq22}
\end{equation}%
where the exponential expansion $e^{i\int A_{\mu }\left( x\right) dx^{\mu
}}=1+i\int A_{\mu }\left( x\right) dx^{\mu }$ applies. In the light of the
Stokes' law one has%
\begin{eqnarray}
&&U_{+}\left( 1,2\right) -U_{-}\left( 1,2\right)=U\left(
1,A\right) \left( i\int_{\partial \fbox{{\tiny AEFBDC}}}A_{\mu
}\left( x\right) dx^{\mu }\right) U\left( B,2\right)  \notag \\
&=&U\left( 1,A\right) \left( i\int_{\fbox{{\tiny AEFBDC}}}\frac{1}{2}F_{\mu
\nu }\left( x\right) dx^{\mu }\wedge dx^{\nu }\right) U\left( B,2\right) ,
\label{Eq3}
\end{eqnarray}%
where $\partial \fbox{{\tiny AEFBDC}}$ is the boundary of the tiny area $%
\fbox{{\tiny AEFBDC}}$ at $x_{0}$. In (\ref{Eq3}) the curvature $F_{\mu \nu
}\left( x\right) $ is the $SU\left( M|N\right) $ gauge field tensor which
has the expansion $F_{\mu \nu }\left( x\right) =F_{\mu \nu }^{ab}\left(
x\right) \hat{e}_{ab}$.

Thus the difference between the path integrals $\left\langle W\left(
L_{-}\right) \right\rangle $ and $\left\langle W\left( L_{+}\right)
\right\rangle $ is%
\begin{eqnarray}
&&\left\langle W\left( L_{+}\right) \right\rangle -\left\langle W\left(
L_{-}\right) \right\rangle  \notag\\
&=&Z^{-1}\int_{\fbox{{\tiny AEFBDC}}}\frac{1}{2}dx^{\mu }\wedge dx^{\nu
}\int \mathcal{D}Ae^{iS}  \notag \\
&&Str\left[ \cdots U\left( 1,A\right) iF_{\mu \nu }^{ab}\left( x\right) \hat{%
e}_{ab}U\left( B,2\right) \cdots U\left( 3,4\right) \cdots \right] .
\end{eqnarray}%
Using the property of the Chern-Simons action (\ref{CSgoodprop}),
one has%
\begin{eqnarray}
&&\left\langle W\left( L_{+}\right) \right\rangle -\left\langle W\left(
L_{-}\right) \right\rangle   \notag \\
&=&\frac{2\pi }{k}Z^{-1}\int_{\fbox{{\tiny AEFBDC}}}d\Sigma ^{\rho }\int
\mathcal{D}A  \notag \\
&&Str\left[ \cdots U\left( 1,A\right) \left( -1\right) ^{\left[ a\right] }%
\hat{e}_{ba}\frac{\partial e^{iS}}{\partial A_{\rho }^{ab}\left( x\right) }%
U\left( B,2\right) \cdots U\left( 3,4\right) \cdots \right]   \notag \\
&=&-\frac{2\pi }{k}Z^{-1}\int_{\fbox{{\tiny AEFBDC}}}d\Sigma ^{\rho }\int
\mathcal{D}Ae^{iS}  \notag \\
&&Str\left[ \cdots U\left( 1,A\right) \left( -1\right) ^{\left[ a\right] }%
\hat{e}_{ba}U\left( B,2\right) \frac{\partial }{\partial A_{\rho
}^{ab}\left( x\right) }\left[ \cdots U\left( 3,4\right) \cdots \right] %
\right] ,  \label{Eq8'}
\end{eqnarray}%
where $d\Sigma ^{\rho }=\frac{1}{2}\epsilon ^{\rho \mu \nu }dx^{\mu }\wedge
dx^{\nu }$ is the surface element of $\fbox{{\tiny AEFBDC}}$, and the
technique of integration by parts has been used. In (\ref{Eq8'}) the propagators $%
\left[ \cdots U\left( 1,A\right) \hat{e}_{ba}U\left( B,2\right)
\right] $ are taken out of the derivative $\frac{\partial
}{\partial A_{\rho }^{ab}\left( x\right) }$ because they are not
impacted by the move of Figure 2. In the remaining propagation
processes $\left[ \cdots U\left( 3,4\right) \cdots \right] $, only
$\left( 3\rightarrow 4\right) $ passes the point $x_{0}$, hence
only $U\left( 3,4\right) $ is impacted by the move. Therefore,
\begin{eqnarray}
&&\left\langle W\left( L_{+}\right) \right\rangle -\left\langle W\left(
L_{-}\right) \right\rangle   \notag \\
&=&-\frac{2\pi }{k}Z^{-1}\int_{\fbox{{\tiny AEFBDC}}}d\Sigma ^{\rho }\int
\mathcal{D}Ae^{iS}\cdot   \notag \\
&&Str\left[ \cdots U\left( 1,A\right) \left( -1\right) ^{\left[ a\right] }%
\hat{e}_{ba}U\left( B,2\right) \cdots \left( \frac{\partial }{\partial
A_{\rho }^{ab}\left( x\right) }U\left( 3,4\right) \right) \cdots \right] .
\label{Eq9}
\end{eqnarray}

Let us examine the $\left( \frac{\partial }{\partial A_{\rho
}^{ab}\left( x\right) }U\left( 3,4\right) \right) $ in
(\ref{Eq9}). It is shown in Figure 2 that
\begin{equation}
U\left( 3,4\right) =e^{i\int_{3}^{4}A_{\lambda }\left( y\right)
dy^{\lambda }}=U\left(3,G\right)e^{\int_{G}^{H}iA_{\lambda
}^{kl}\left( y\right) \hat{e}_{kl}dy^{\lambda }}U\left(H,4\right),
\label{dylambda}
\end{equation}%
where $\overline{\mathrm{GH}}$ is a short segment passing $x_{0}$.
Thus
\begin{equation}
\frac{\partial }{\partial A_{\rho }^{ab}\left( x\right) }U\left( 3,4\right)
=U\left( 3,G\right) \left[ \int_{G}^{H}i\delta ^{3}\left( x-x_{0}\right)
dx^{\rho }\hat{e}_{ab}e^{\int_{G}^{H}iA_{\lambda }^{kl}\left( y\right) \hat{e%
}_{kl}dy^{\lambda }}\right] U\left( H,4\right) ,
\end{equation}%
and (\ref{Eq9}) becomes%
\begin{eqnarray}
&&\left\langle W\left( L_{+}\right) \right\rangle -\left\langle W\left(
L_{-}\right) \right\rangle  \notag \\
&=&-i\frac{2\pi }{k}Z^{-1}\int_{\fbox{{\tiny AEFBDC}}}\int_{G}^{H}\delta
^{3}\left( x-x_{0}\right) d\Sigma ^{\rho }\otimes dx^{\rho }\int \mathcal{D}%
Ae^{iS}\cdot  \notag \\
&&Str\left[ \cdots U\left( 1,A\right) \left( -1\right) ^{\left[ a\right] }%
\hat{e}_{ba}U\left( B,2\right) \cdots U\left( 3,x_{0}\right) \hat{e}%
_{ab}U\left( x_{0},4\right) \cdots \right] ,  \label{Eq10}
\end{eqnarray}%
where the $dx^{\rho }$ is along the direction of the segment $\overline{%
\mathrm{GH}}$. In (\ref{Eq10}) a volume integral is recognized:
\begin{equation}
\left[ \text{vol}\right] _{x_{0}}=\int_{\fbox{{\tiny AEFBDC}}%
}\int_{G}^{H}\delta ^{3}\left( x-x_{0}\right) d\Sigma ^{\rho }\otimes
dx^{\rho },  \label{tinyvol}
\end{equation}
which has the evaluation
\begin{equation}
\left[ \text{vol}\right] _{x_{0}}\left\{
\begin{array}{ll}
=0, & \text{trivial;} \\
=\pm 1, & \text{non-trivial.}%
\end{array}
\right.  \label{volevalue}
\end{equation}
In detail,

\begin{itemize}
\item $\left[ \text{vol}\right] _{x_{0}}=0$ describes the trivial
case that in Figure 2 the $dx^{\rho }$ is parallel to the plane of
$\fbox{{\tiny AEFBDC}}$; namely, the move from
$\overline{\mathrm{ACDB}}$ to $\overline{\mathrm{AEFB}}$ is done
by sliding along $\overline{\mathrm{3GH4}}$. Therefore $d\Sigma
^{\rho }\otimes dx^{\rho}=0$.

\item $\left[ \text{vol}\right] _{x_{0}}=1$ describes the
non-trivial move $L_{-}\rightarrow $ $L_{+}$, where $dx^{\rho }$
is perpendicular to $\fbox{{\tiny AEFBDC}}$ and $d\Sigma ^{\rho
}\otimes dx^{\rho}=1$; otherwise, $\left[ \text{vol}\right]
_{x_{0}}=-1$ for $L_{+}\rightarrow $ $L_{-}$, where $dx^{\rho }$
is perpendicular to $\fbox{{\tiny AEFBDC}}$ but $d\Sigma ^{\rho
}\otimes dx^{\rho}=-1$. The case we come across in Figure 2 is the
former, so $\left[ \text{vol}\right] _{x_{0}}=1$.
\end{itemize}

\noindent Therefore, (\ref{Eq10}) becomes
\begin{eqnarray}
&&\left\langle W\left( L_{+}\right) \right\rangle -\left\langle W\left(
L_{-}\right) \right\rangle  \notag \\
&=&-i\frac{2\pi }{k}Z^{-1}\int \mathcal{D}Ae^{iS}  \notag \\
&&Str\left[ \cdots U\left( 1,A\right) \left( -1\right) ^{\left[ b\right] }%
\hat{e}_{ab}U\left( B,2\right) \cdots U\left( 3,x_{0}\right) \hat{e}%
_{ba}U\left( x_{0},4\right) \cdots \right] .  \label{Eq14}
\end{eqnarray}

\section{Skein Relations}

In this section the $S_{L}\left( \alpha ,\beta ,z\right) $
polynomial invariant for knots in the $SU\left( M|N\right) $ CS
field theory will be derived from (\ref{Eq14}), and its
relationship to the HOMFLY and Jones polynomials will be
discussed.

Under the fundamental representation the entries of the matrices $\hat{e}%
_{ab}$ satisfy the Fierz identity \cite{FierzId}%
\begin{equation}
\left( -1\right) ^{\left[ b\right] }\left( \hat{e}_{ab}\right) _{ij}\left(
\hat{e}_{ba}\right) _{kl}=\left( -1\right) ^{\left[ j\right] }\delta
_{il}\delta _{jk}-\frac{1}{M-N}\delta _{ij}\delta _{kl}.  \label{FierzId}
\end{equation}%
Hence (\ref{Eq14}) leads to%
\begin{eqnarray}
&&\left\langle W\left( L_{+}\right) \right\rangle -\left\langle W\left(
L_{-}\right) \right\rangle  \notag \\
&=&-i\frac{2\pi }{k}Z^{-1}\int \mathcal{D}Ae^{iS} \cdot
\notag\\
&&Str\left[ \cdots U\left( 1,A\right) U\left( x_{0},4\right)
\cdots \right]
Str\left[ U\left( B,2\right) \cdots U\left( 3,x_{0}\right) \right]  \notag \\
&&+i\frac{2\pi }{k}\frac{1}{M-N}Z^{-1}\int \mathcal{D}Ae^{iS}\cdot  \notag \\
&&Str\left[ \cdots U\left( 1,A\right) U\left( B,2\right) \cdots U\left(
3,x_{0}\right) U\left( x_{0},4\right) \cdots \right] .  \label{Eq15}
\end{eqnarray}%
When the points $A$ and $B$ approaching $x_{0}$, the first term of
(\ref{Eq15}) corresponds to the non-crossing case $L_{0}$ in
Figure 1(c). For the second term, however, one has two ways to
connect $A$ and $B$ --- the undercrossing and the overcrossing
--- in order to form a
propagation process $\left( 1\rightarrow 2\right) $. Treating
these two crossing ways equally, one has
\begin{equation}
\left( 1-i\frac{\pi }{k}\frac{1}{\left( M-N\right) }\right) \left\langle
W\left( L_{+}\right) \right\rangle -\left( 1+i\frac{\pi }{k}\frac{1}{\left(
M-N\right) }\right) \left\langle W\left( L_{-}\right) \right\rangle =-i\frac{%
2\pi }{k}\left\langle W\left( L_{0}\right) \right\rangle .  \label{beta1}
\end{equation}
Then, considering the weak coupling limit of large $k$
\cite{WittenCMP1989knots},
we define%
\begin{equation}
\beta =1-i\frac{\pi }{k}\frac{1}{\left( M-N\right) }+O\left( \frac{1}{k^{2}}%
\right) ,\ \ \ \ \ z=-i\frac{2\pi }{k}+O\left( \frac{1}{k^{2}}\right) ,
\label{beta2}
\end{equation}%
and obtain an important skein relation%
\begin{equation}
\beta \left\langle W\left( L_{+}\right) \right\rangle -\beta
^{-1}\left\langle W\left( L_{-}\right) \right\rangle =z\left\langle W\left(
L_{0}\right) \right\rangle .  \label{beta3}
\end{equation}

For the purpose of examining knot writhing, let us consider the
special case that the point $x_{2}$ is identical to $x_{3}$ in
Figure 1. Then in (\ref{Eq14}) one has
\begin{equation}
\lim_{B\rightarrow x_{0};x_{2}=x_{3}}U\left( B,2\right) \cdots U\left(
3,x_{0}\right) =I,  \label{alpha1}
\end{equation}%
and%
\begin{eqnarray}
&&\left\langle W\left( \hat{L}_{+}\right) \right\rangle -\left\langle
W\left( \hat{L}_{-}\right) \right\rangle  \notag \\
&=&-i\frac{2\pi }{k}Z^{-1}\int \mathcal{D}Ae^{iS}Str\left[ \cdots U\left(
1,A\right) \left( -1\right) ^{\left[ b\right] }\hat{e}_{ab}\hat{e}%
_{ba}U\left( x_{0},4\right) \cdots \right] ,  \label{Eq14'}
\end{eqnarray}%
where $\hat{L}_{+}$ and $\hat{L}_{-}$ are two writhing situations
shown in Figure 3(a) and 3(b). Figure 3(c) shows the non-writhing
situation $\hat{L}_{0}$.
\begin{figure}[ht]
\centering \includegraphics[scale=0.7]{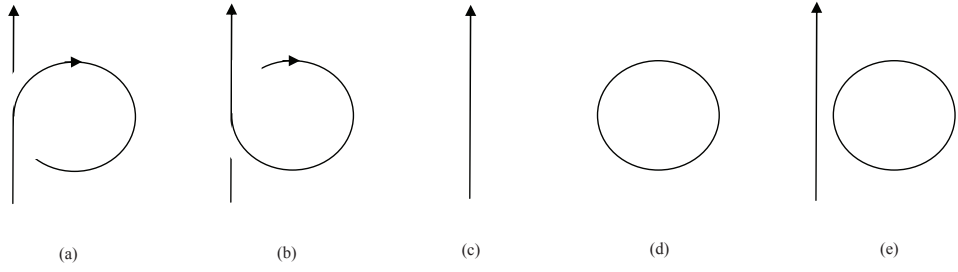}
\caption{Typical Configurations: (a) writhing $\hat{L}_{+}$; (b)
writhing $\hat{L}_{-}$;
 (c) non-writhing $\hat{L}_{0}$; (d) trivial circle $\hat{L}_{c}$;
 (e) non-intersecting union $\hat{L}_{i}$.}
\end{figure}

In the above the factor $\left( -1\right) ^{\left[ b\right] }\hat{e}_{ab}%
\hat{e}_{ba}$ is the Casimir operator%
\begin{equation}
\left( -1\right) ^{\left[ b\right] }\hat{e}_{ab}\hat{e}_{ba}=2C_{2}I,\ \ \ \
\ \ \ \ C_{2}=\frac{\left( M-N\right) ^{2}-1}{2\left( M-N\right) },\ \ M\neq
N.  \label{Cas2-1}
\end{equation}%
When $A$ approaches $x_{0}$ one has%
\begin{equation}
\left\langle W\left( \hat{L}_{+}\right) \right\rangle -\left\langle W\left(
\hat{L}_{-}\right) \right\rangle =-i\frac{4\pi }{k}C_{2}\left\langle W\left(
\hat{L}_{0}\right) \right\rangle ,  \label{alpha0}
\end{equation}%
where $\left\langle W\left( \hat{L}%
_{0}\right) \right\rangle =Z^{-1}\int \mathcal{D}Ae^{iS}Str\left[ \cdots
U\left( 1,x_{0}\right) U\left( x_{0},4\right) \cdots \right] $. The move $%
\hat{L}_{-}\rightarrow $ $\hat{L}_{+}$ is a change of the writhe
of the path segment. In this regard an intermediate stage
$\hat{L}_{0}$ can be inserted and the move becomes
$\hat{L}_{-}\rightarrow $
$\hat{L}_{0}\rightarrow $ $%
\hat{L}_{+}$. Then the correlation function becomes $\left\langle
W\left( \hat{L}_{+}\right) \right\rangle -\left\langle W\left(
\hat{L}_{-}\right) \right\rangle =\left[ \left\langle W\left( \hat{L}%
_{+}\right) \right\rangle -\left\langle W\left( \hat{L}_{0}\right)
\right\rangle \right] +\left[ \left\langle W\left( \hat{L}_{0}\right)
\right\rangle -\left\langle W\left( \hat{L}_{-}\right) \right\rangle \right]
.$ The two subprocesses $\hat{L}_{-}\rightarrow \hat{L}_{0}$ and $\hat{L}%
_{0}\rightarrow $ $\hat{L}_{+}$ should be equivalent, hence\\
$\left\langle W\left( \hat{L}_{+}\right) \right\rangle
-\left\langle W\left( \hat{L}_{0}\right) \right\rangle
=\left\langle W\left( \hat{L}_{0}\right) \right\rangle
-\left\langle W\left( \hat{L}_{-}\right) \right\rangle
=-i\frac{2\pi }{k} C_{2}\left\langle W\left( \hat{L}_{0}\right)
\right\rangle$, and we arrive at another skein relation
\begin{equation}
\left\langle W\left( \hat{L}_{+}\right) \right\rangle =\alpha \left\langle
W\left( \hat{L}_{0}\right) \right\rangle ,\ \ \left\langle W\left( \hat{L%
}_{-}\right) \right\rangle =\alpha ^{-1}\left\langle W\left( \hat{L}%
_{0}\right) \right\rangle ,\ \ \alpha =1-i\frac{2\pi
}{k}C_{2}+O\left( \frac{1}{k^{2}}\right) .  \label{alpha4}
\end{equation}

Besides (\ref{beta3}) and (\ref{alpha4}), one needs the
correlation function for the trivial circle $\hat{L}_{c}$ shown in
Figure 3(d):
\begin{equation}
\left\langle W\left( \hat{L}_{c}\right) \right\rangle =Z^{-1}\int \mathcal{D}%
Ae^{iS}Str\left[ \hat{L}_{c}\right] =Z^{-1}\int \mathcal{D}Ae^{iS}Str\left[ I%
\right] =\left( M-N\right) .
\end{equation}%
Thus, in summary, we have acquired the following skein relations
for knots in the $SU\left(M|N\right) $ CS field theory:%
\begin{eqnarray}
&&\left\langle W\left( \hat{L}_{c}\right) \right\rangle =M-N\ \ \
\ \ \ \ \ \ \ \ \ \ \ \left(
M\neq N\right) ,  \label{SLpoly0} \\
&&\left\langle W\left( \hat{L}_{+}\right) \right\rangle =\alpha \left\langle
W\left( \hat{L}_{0}\right) \right\rangle ,\ \ \ \ \left\langle W\left( \hat{L}%
_{-}\right) \right\rangle =\alpha ^{-1}\left\langle W\left( \hat{L}%
_{0}\right) \right\rangle ,  \label{SLpoly2} \\
&&\beta \left\langle W\left( L_{+}\right) \right\rangle -\beta
^{-1}\left\langle W\left( L_{-}\right) \right\rangle =z\left\langle W\left(
L_{0}\right) \right\rangle ,  \label{SLpoly3}
\end{eqnarray}%
with
\begin{equation}
\alpha =1-i\frac{2\pi }{k}C_{2}+O\left( \frac{1}{k^{2}}\right) ,\ \ \ \beta
=1-i\frac{\pi }{k}\frac{1}{\left( M-N\right) }+O\left( \frac{1}{k^{2}}%
\right) ,\ \ \ z=-i\frac{2\pi }{k}+O\left( \frac{1}{k^{2}}\right)
. \label{coefficients}
\end{equation}
These relations present a polynomial invariant $\left\langle W\left(
L\right) \right\rangle $ for the knots, known as the $S_{L}\left( \alpha
,\beta ,z\right) $ polynomial proposed by Guadagnini et al. \cite%
{GuadagniniNPB1990,GuadagniniBook}.

It is checked that Eq.(\ref{SLpoly2}) is consistent with
(\ref{SLpoly3}). Considering the special case $x_{2}=x_{3}$ for
(\ref{beta3}) there is
\begin{equation}
\beta \left\langle W\left( \hat{L}_{+}\right) \right\rangle -\beta
^{-1}\left\langle W\left( \hat{L}_{-}\right) \right\rangle =z\left\langle
W\left( \hat{L}_{i}\right) \right\rangle ,  \label{alpha3}
\end{equation}%
where $\hat{L}_{i}$ is the non-intersecting union of a trivial
circle and a line segment shown in Figure 3(e). The LHS of
(\ref{alpha3}) gives $\beta \left\langle W\left(
\hat{L}_{+}\right) \right\rangle -\beta ^{-1}\left\langle W\left(
\hat{L}_{-}\right) \right\rangle =\left( \beta
\alpha -\beta ^{-1}\alpha ^{-1}\right) \left\langle W\left( \hat{L}%
_{0}\right) \right\rangle $ with respect to (\ref{SLpoly2}). The RHS of (\ref%
{alpha3}) is
\begin{equation}
z\left\langle W\left( \hat{L}_{i}\right) \right\rangle
=zZ^{-1}\int \mathcal{D}Ae^{iS}Str\left[ \cdots U\left( 1,4\right) \cdots %
\right] Str\left[ \hat{L}_{c}\right] =z\left( M-N\right)
\left\langle W\left( \hat{L}_{0}\right) \right\rangle .
\end{equation}
Hence $\beta \alpha -\beta ^{-1}\alpha
^{-1}=z\left( M-N\right) $, which is consistent with the definitions of $%
\alpha ,\ \beta $ and $z$.

The $S_{L}\left( \alpha ,\beta ,z\right) $ polynomial is
regular-isotopic, but not ambient-isotopic. Namely, $\langle
W\left(L\right)\rangle$ is invariant under the type-II and -III
Reidemeister moves (shown in Figure 4), but is not invariant under
the type-I move. Indeed,

\begin{itemize}
\item in a type-II move, path variation of Figure 2 takes place at
both the points $x_{0a}$ and $x_{0b}$. Then there are volumes of
variation given in (\ref{tinyvol}) at both $x_{0a}$ and $x_{0b}$,
which are marked as $\left[ \text{vol}\right] _{x_{0a}}$ and
$\left[ \text{vol}\right] _{x_{0b}}$ respectively. It can be
checked that $\left[ \text{vol}\right] _{x_{0a}}$ and $\left[
\text{vol}\right] _{x_{0b}}$ take
opposite sign: $\left[ \text{vol}\right] _{x_{0a}}=1,\ \left[ \text{vol}%
\right] _{x_{0b}}=-1$. Hence totally the type-II move causes no
variation in the correlation function;

\item in a type-III move, there are neither \textquotedblleft
undercrossing to overcrossing\textquotedblright nor
\textquotedblleft overcrossing to undercrossing \textquotedblright
moves taking place, so the volume of variation is zero, and the
type-III move causes no variation in the correlation function;

\item in a type-I move, the variation of the correlation function
is given by (\ref{SLpoly2}).
\end{itemize}

\begin{figure}[th]
\centering \includegraphics[scale=0.9]{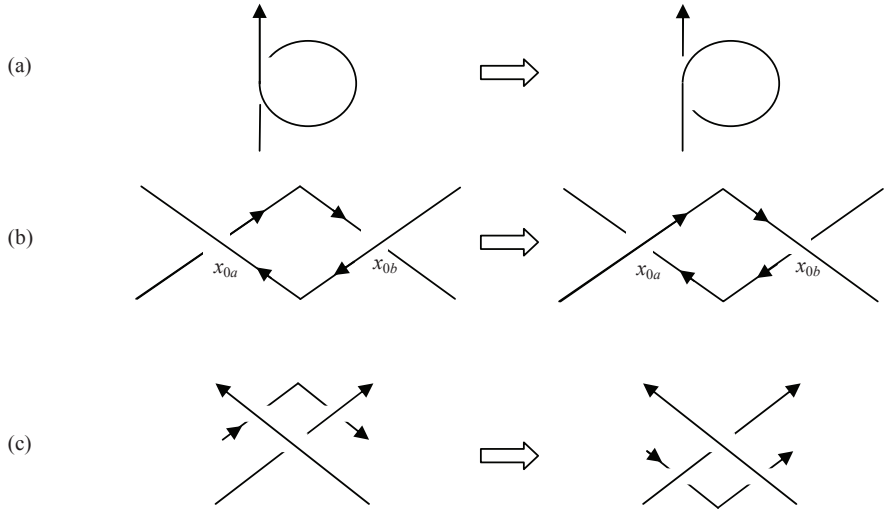}
\caption{Reidemeister Moves: (a) Type-I; (b) Type-II; (c)
Type-III.}
\end{figure}

\noindent In following the relationships between the $S_{L}\left(
\alpha ,\beta ,z\right) $ polynomial and other knot polynomial
invariants will be studied. $\left\langle W\left( L\right)
\right\rangle $ will be modified to be an ambient-isotopic
invariant, and a difference between the normal and super Lie gauge
groups, $SU\left( N\right) $ and $SU\left( M|N\right) $, will
arise from the Jones polynomial.

Firstly, the ambient-isotopic HOMFLY knot polynomial invariant can be
constructed from $\left\langle W\left( L\right) \right\rangle $ by
introducing a factor describing knot writhing:%
\begin{equation}
\left\langle P\left( L\right) \right\rangle =\alpha ^{-\omega \left(
L\right) }\left\langle W\left( L\right) \right\rangle .  \label{defHOMFLY}
\end{equation}%
Here $\omega \left( L\right) $ is the writhe number of a knot $L$, defined as%
\begin{equation}
\omega \left( L_{\pm }\right) =\omega \left( L_{0}\right) +\epsilon \left(
L_{\pm };x_{0}\right) =\omega \left( L_{0}\right) \pm 1,  \label{writhnum1}
\end{equation}%
where $\epsilon \left( L_{\pm };x_{0}\right) $ is the sign of the
crossing point $x_{0}$ on $L_{\pm }$: $\epsilon \left( L_{\pm
};x_{0}\right) =\pm 1$. For $\hat{L}_{+},\ \hat{L}_{-}$ and $\hat{L}_{0}$, (%
\ref{writhnum1}) reads%
\begin{equation}
\omega \left( \hat{L}_{+}\right) =\omega \left( \hat{L}_{0}\right)
+1,\ \ \ \ \omega \left( \hat{L}_{-}\right) =\omega \left(
\hat{L}_{0}\right) -1. \label{writhnum2}
\end{equation}
(\ref{writhnum2}) means that $\hat{L}_{+}$ contributes a $1$\ to
the writhe number, while $\hat{L}_{-}$ contributes a $\left(
-1\right) $. Then using (\ref{SLpoly2}) and (\ref{defHOMFLY}) one
has
\begin{equation}
\left\langle P\left( \hat{L}_{+}\right) \right\rangle =\left\langle P\left(
\hat{L}_{0}\right) \right\rangle ,\ \ \ \ \left\langle P\left( \hat{L}%
_{-}\right) \right\rangle =\left\langle P\left( \hat{L}_{0}\right)
\right\rangle ,
\end{equation}
meaning $\left\langle P\left( L\right) \right\rangle $ is
invariant under the type-I Reidemeister move. Furthermore $\langle
P\left(L\right)\rangle$ satisfies
\begin{equation}
\left( \alpha \beta \right) \left\langle P\left( L_{+}\right) \right\rangle
-\left( \alpha \beta \right) ^{-1}\left\langle P\left( L_{-}\right)
\right\rangle =zP\left( L_{0}\right) .
\end{equation}
Hence one arrives at the skein relations for $\left\langle P\left(
L\right)
\right\rangle :$%
\begin{eqnarray}
&&\left\langle P\left( \hat{L}_{c}\right) \right\rangle =M-N,
\label{HOMFLY0} \\
&&t\left\langle P\left( L_{+}\right) \right\rangle -t^{-1}\left\langle
P\left( L_{-}\right) \right\rangle =z\left\langle P\left( L_{0}\right)
\right\rangle ,  \label{HOMFLY1}
\end{eqnarray}%
where%
\begin{equation}
t\equiv \alpha \beta =1-i\frac{2\pi }{k}\frac{\left( M-N\right) }{2}+O\left(
\frac{1}{k^{2}}\right) ,\ \ \ \ \ \ z=-i\frac{2\pi }{k}+O\left( \frac{1}{%
k^{2}}\right) .  \label{HOMFLY2}
\end{equation}
(\ref{HOMFLY0}) can be obtained from (\ref{HOMFLY1}) by considering $%
t\left\langle P\left( \tilde{L}_{+}\right) \right\rangle -t^{-1}\left\langle
P\left( \tilde{L}_{-}\right) \right\rangle =z\left\langle P\left\langle
\tilde{L}_{c}\right\rangle \right\rangle $, where $\tilde{L}_{+},\ \tilde{L}%
_{-}$ and $\tilde{L}_{c}$ denote unknots shown in Figure 5.
\begin{figure}[th]
\centering \includegraphics[scale=0.8]{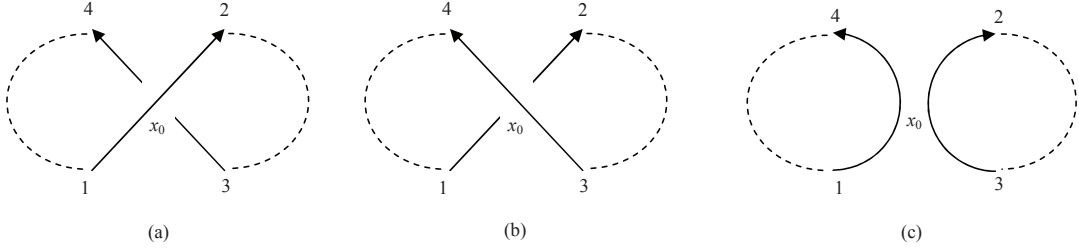}
\caption{Unknots: (a) $\tilde{L}_{+}$; (b) $\tilde{L}_{-}$; (c)
$\tilde{L}_{c}$.}
\end{figure}

\noindent Eqs.(\ref{HOMFLY0}) and (\ref{HOMFLY1}) show
$\left\langle P\left( L\right) \right\rangle $ is an
ambient-isotopic HOMFLY polynomial invariant.

Secondly, if specially $M-N=2$ in (\ref{HOMFLY0}) to
(\ref{HOMFLY2}), the $z$ is related to $t$ as
$z=t^{\frac{1}{2}}-t^{-\frac{1}{2}}$, up to the first order. This
means that in the $SU\left( N+2|N\right) $ CS field
theory, under the fundamental representation there is a knot polynomial $%
\left\langle V\left( L\right) \right\rangle \equiv \left\langle P\left(
L\right) \right\rangle $ which satisfies the skein relation%
\begin{equation}
t\left\langle V\left( L_{+}\right) \right\rangle -t^{-1}\left\langle V\left(
L_{-}\right) \right\rangle =\left( t^{\frac{1}{2}}-t^{-\frac{1}{2}}\right)
\left\langle V\left( L_{0}\right) \right\rangle .  \label{JonesPoly1}
\end{equation}%
This $\left\langle V\left( L\right) \right\rangle $ is known as
the Jones polynomial \footnote{\normalsize Compared to the
standard conventions adopted in
mathematics, there is a sign difference in the skein relation (\ref%
{JonesPoly1}) of the Jones polynomial. See \cite{GuadagniniBook}
for this discussion.}. Therefore there are a series of CS theories
with Lie super gauge group $SU\left( N+2|N\right),\ N\in
\mathbb{Z}^{+},$ which have the Jones polynomial. This is
different from the situation of the CS theory with normal Lie
group $SU\left( N\right)$ --- it is known that under the
fundamental representation, only the $SU\left( 2\right)$ theory
has the Jones polynomial invariant among all $SU\left( N\right)$
CS theories, $\ N=2,3,\cdots \ $
\cite{WittenCMP1989knots,GuadagniniBook,Labastida}.

Different choices of gauge groups with different algebraic
representations lead to different knot polynomials in CS field
theories \cite{Labastida}. In our further work the relationship
between the $S_{L}\left( \alpha ,\beta ,z\right) $ and the
Kauffman polynomials in the $OSp\left( 1|2\right) $ CS field
theory will be studied.

Finally, the $\alpha ,\ \beta $ and $z$ in the $S_{L}\left( \alpha
,\beta ,z\right) $ polynomial and the $t$ in the HOMFLY polynomial
can be
expressed in a unified way. Introducing a variable%
\begin{equation}
q=e^{-i\frac{2\pi }{k}},
\end{equation}%
$\alpha ,\ \beta $, $z$ and $t$ can be regarded as the lower order
expansions of the $q$ exponentials \cite%
{GuadagniniNPB1990,GuadagniniBook,WittenCMP1989knots}: $\alpha =q^{C_{2}}=q^{%
\frac{\left( M-N\right) ^{2}-1}{2\left( M-N\right) }},\ \beta =q^{\frac{1}{%
2\left( M-N\right) }},\ z=q^{\frac{1}{2}}-q^{-\frac{1}{2}}$ and $t=q^{\frac{%
M-N}{2}}.$ Then the $S_{L}\left( \alpha ,\beta ,z\right) $ shown
in (\ref{SLpoly0})--(\ref{SLpoly3}) and HOMFLY polynomial in
(\ref{HOMFLY0})--(\ref{HOMFLY1}) can be written more elegantly as
\begin{eqnarray}
&&\left\langle W\left( \hat{L}_{c}\right) \right\rangle =M-N\ \ \ \ \left(
M\neq N\right) , \\
&&\left\langle W\left( \hat{L}_{+}\right) \right\rangle =q^{\frac{\left(
M-N\right) ^{2}-1}{2\left( M-N\right) }}\left\langle W\left( \hat{L}%
_{0}\right) \right\rangle , \\
&&\left\langle W\left( \hat{L}_{-}\right) \right\rangle =q^{-\frac{\left(
M-N\right) ^{2}-1}{2\left( M-N\right) }}\left\langle W\left( \hat{L}%
_{0}\right) \right\rangle , \\
&&q^{\frac{1}{2\left( M-N\right) }}\left\langle W\left( L_{+}\right)
\right\rangle -q^{-\frac{1}{2\left( M-N\right) }}\left\langle W\left(
L_{-}\right) \right\rangle =\left(q^{\frac{1}{2}}-q^{-\frac{1}{2}}\right)\left\langle
W\left( L_{0}\right) \right\rangle ,
\end{eqnarray}%
and%
\begin{eqnarray}
&&\left\langle P\left( \hat{L}_{c}\right) \right\rangle =\frac{q^{\frac{M-N}{%
2}}-q^{-\frac{M-N}{2}}}{q^{\frac{1}{2}}-q^{-\frac{1}{2}}}, \\
&&q^{\frac{M-N}{2}}\left\langle P\left( L_{+}\right) \right\rangle -q^{-%
\frac{M-N}{2}}\left\langle P\left( L_{-}\right) \right\rangle =\left( q^{%
\frac{1}{2}}-q^{-\frac{1}{2}}\right) \left\langle P\left( L_{0}\right)
\right\rangle .
\end{eqnarray}

\section{Conclusion}

In this paper we have studied knots in the CS field theory with
gauge group $SU\left( M|N\right) $. In Section 2, the notation
for the fundamental representation of the Lie superalgebra $\mathfrak{su}%
\left( M|N\right) $ is fixed, and an important property of the CS
action, Eq.(\ref{CSgoodprop}), is presented. In Section 3,
variation of the correlation function of Wilson loops is
rigorously studied. In Section 4, the variation of correlation
functions (\ref{Eq14}) is discussed for different link
configurations. It is addressed that the path integrals have been
formally expressed as propagators instead of being integrated out.
A rigorous development of techniques for path integrals awaits
future advances in the mathematical theory of functional
integrals. From the formal analysis the $S_{L}\left( \alpha ,\beta
,z\right) $ knot polynomial and its skein relations,
(\ref{SLpoly0}) to (\ref{SLpoly3}), are obtained. In terms of the
$S_{L}\left( \alpha ,\beta ,z\right) $ polynomial the HOMFLY and
Jones knot polynomials as well as their skein relations
(\ref{HOMFLY0}) to (\ref{JonesPoly1}) have been derived by
considering the knot writhing.

\section{Acknowledgment}

The author is indebted to Prof. R.B. Zhang and Dr. W.L. Yang for instructive
advices and warmhearted help. This work was financially supported by the
USYD Postdoctoral Fellowship of the University of Sydney, Australia.


\begin{thebibliography}{99}
\bibitem{GuadagniniBook} Guadagnini, E.: \textit{The Link Invariants of the
Chern-Simons Field Theory}, Walter de Gruyter \& Co., Berlin, 1993.

\bibitem{JackiwWZW} Deser, S., Jackiw, R., Templeton, S.: \textit{Ann. Phys.}
\textbf{140} (1982) 372;\newline Jackiw, R.: \textit{Topological
Investigations of Quantized Gauge Theories}, in \textit{Current
Algebras and Anomalies}, edited by Treiman S.B., Jackiw R., Zumino
B. and Witten E., World Scientific, 1985.

\bibitem{WittenCMP1989knots} Witten, E.: \textit{Commun. Math. Phys.}
\textbf{121} (1989) 351.

\bibitem{GuadagniniNPB1990} Guadagnini, E., Martellini, M., Mintchev, M.:
\textit{Nucl. Phys. B} \textbf{330} (1990) 575;\newline
Cotta-Ramusino, P., et al.: \textit{%
Nucl. Phys. B} \textbf{330} (1990) 557.

\bibitem{Kauffmanbook} Kauffman L.H.: \textit{Knots and Physics}, 2nd ed.,
World Scientific, Singapore, 2001;\newline
Kauffman L.H.: \textit{Rep. Prog. Phys.} \textbf{68} (2005) 2829.

\bibitem{Bar-Natan} Bar-Natan, D.: \textit{Topology} \textbf{34} (1995) 423.

\bibitem{BirmanLin} Birman, J.S., Lin, X.S.: \textit{Invent. Math.} \textbf{%
111} (1993) 225.

\bibitem{Labastida} Labastida, J.M.F.: \textit{Chern-Simons Gauge Theory: Ten Years After}, in
\textit{Trends in Theoretical Physics II}, H. Falomir, R. Gamboa, F.
Schaposnik, eds., American Institute of Physics, New York, 1999, CP 484
(1-41), available at: hep-th/9905057.

\bibitem{Marino2005RMP} Marino, M., \textit{Rev. Mod. Phys.} \textbf{77}
(2005) 675.

\bibitem{2+1quantumgravity} Gambini, R., Pullin, J.: \textit{Loops, Knots,
Gauge Theories and Quantum Gravity}, Cambridge, Cambridge
University Press, 1996;\newline Li, W., Song, W., Strominger, A.:
\textit{JHEP} \textbf{0804} (2008) 082.

\bibitem{LieSuperalgebra} Mudry, C., Chamon, C., Wen, X.-G.: \textit{Nucl.
Phys. B} \textbf{466} (1996) 383;\newline
Guruswamy, S., LeClair, A., Ludwig, A.W.W.: \textit{Nucl. Phys. B} \textbf{%
583} (2000) 475;\newline
Weinberg, E.J., Yi, P.: \textit{Phys. Rep.} \textbf{438} (2007) 65.

\bibitem{Scheunert} Scheunert, M., Nahm, W., Rittenberg, V.: \textit{J.
Math. Phys.} \textbf{18} (1977) 146.

\bibitem{RBZetalAlgeRep} Gould, M., Zhang, R.: \textit{J. Math. Phys.} \textbf{31}
(1990) 1524; \textit{ibid.}, \textit{J. Math. Phys.} \textbf{31} (1990) 2552.%
\newline
Scheunert, M.; Zhang, R.: \textit{J. Algebra} \textbf{292} (2005) 324.

\bibitem{RBZlinkInv} Zhang, R., Gould, M., Bracken, A.: \textit{Commun.
Math. Phys.} \textbf{137} (1991) 13;\newline
Kauffman, L.H., Saleur, H.: \textit{Commun. Math. Phys.} \textbf{141} (1991)
293;\newline
Gould, M., Tsohantjis, I., Bracken, A.: \textit{Rev. Math. Phys.} \textbf{5}
(1993) 533;\newline
Links, J., Gould, M., Zhang, R.: \textit{Rev. Math. Phys.} \textbf{5} (1993)
345;\newline
Links, J., Zhang, R.: \textit{J. Math. Phys.} \textbf{35} (1994) 1377.

\bibitem{CSsuperalge} Bourdeau, M., et al.: \textit{Nucl. Phys. B} \textbf{%
372} (1992) 303;\newline
Ennes, I.P., et al.: \textit{Int. J. Mod. Phys. A} \textbf{13} (1998) 2931;%
\newline
Gaiotto, D., Witten, E.: \textit{Janus Configurations, Chern-Simons
Couplings, and the }$\theta $\textit{-Angle in }$\mathcal{N}=4$\textit{\
Super Yang-Mills Theory,} available at: arXiv:0804.2907 [hep-th].

\bibitem{LiuX} Duan, Y.S., Liu, X., Fu, L.B.: \textit{Phys. Rev. D} \textbf{67}
(2003) 085022;\newline Duan, Y.S., Liu, X.: \textit{JHEP}
\textbf{02} (2004) 028.

\bibitem{KacAnnMath1977} Kac, V.G.: \textit{Adv. Math.} \textbf{26} (1977) 8.

\bibitem{dictionary} Frappat, L., Sciarrino, A., Sorba, P.: \textit{%
Dictionary on Lie Algebras and Superalgebras}, Academic Press, 2000.

\bibitem{glNNdef-NPB1994} Isidro J.M., Ramallo A.V.: \textit{Nucl. Phys. B}
\textbf{414} (1994) 715.

\bibitem{WLYangJMP} Yang, W.L., Zhang, Y.Z., Liu, X.: \textit{J. Math. Phys.}
\textbf{48} (2007) 053514.

\bibitem{CronstromMickelsson} Cronstr\"{o}m, C., Mickelsson, J.: \textit{J.
Math. Phys.} \textbf{24} (1983) 2528.

\bibitem{FierzId} Bagger, J., Lambert, N.: \textit{Phys. Rev. D} \textbf{77}
(2008) 065008;\newline Figueroa-O¡¯Farrill, J., Hackett-Jones, E.,
Moutsopoulos, G.: \textit{Class. Quant. Grav.} \textbf{24} (2007)
3291.

\end{thebibliography}
\end{document}